\documentclass{llncs}

\usepackage{amsmath}
\usepackage{amsfonts}
\usepackage{amssymb}
\usepackage{graphicx}
\usepackage{color}
\usepackage{algorithmic}
\usepackage{url}
\usepackage[ruled,linesnumbered]{algorithm2e}
\usepackage{epsfig}

\newcommand{\ben}{\begin{enumerate}}
\newcommand{\een}{\end{enumerate}}

\newcommand{\hide}[1]{}

\newcommand{\pegasus}{\textsc{PeGaSus}\xspace}

\newcommand{\field}[1]{\mathbb{#1}} 

\newcommand{\beql}[1]{\begin{equation}\label{#1}}
\newcommand{\eeq}{\end{equation}}
\newcommand{\comment}[1]{}

\newcommand{\Abs}[1]{{\left|{#1}\right|}}

\newcommand{\Mean}[1]{{\mathbb E}\left[{#1}\right]}

\newcommand{\Prob}[1]{{{\bf{Pr}}\left[{#1}\right]}}
\newcommand{\Set}[1]{{\left\{{#1}\right\}}}

\newcommand{\One}[1]{{\mathbf 1}\left(#1\right)}

\begin{document}

\mainmatter    

\title{Approximate Triangle Counting}
\titlerunning{Triangles}

\author{Charalampos E. Tsourakakis\inst{1} \and Mihail N. Kolountzakis \inst{2}
\and Gary L. Miller\inst{1}
}
\authorrunning{Charalampos E. Tsourakakis et al.}   

\tocauthor{Charalampos E. Tsourakakis, Mihail N. Kolountzakis, Gary L. Miller}

\institute{School of Computer Science, Carnegie Mellon University, Pittsburgh PA 15213, USA,\\
\email{ctsourak@cs.cmu.edu}, \\ WWW home page:
\texttt{http://www.cs.cmu.edu/\homedir ctsourak} \\
\email{glmiller@cs.cmu.edu} \\ WWW home page:
\texttt{http://www.cs.cmu.edu/\homedir glmiller}
\and
Department of Mathematics, University of Crete, Knossos Ave., 714 09 Iraklio, Greece,
\email{kolount@math.uoc.gr}, \\\ WWW home page:
\texttt{http://fourier.math.uoc.gr/\homedir mk}
}
\maketitle

\begin{abstract}
Triangle counting is an important problem in graph mining. Clustering coefficients of vertices and the transitivity ratio of the graph
are two metrics often used in complex network analysis. Furthermore, triangles have been used successfully in several real-world applications.
However, exact triangle counting is an expensive computation. 
In this paper we present the analysis of a practical sampling algorithm for counting triangles in graphs. 
Our analysis yields optimal values for the sampling rate, thus resulting in tremendous speedups ranging from \emph{2800}x to \emph{70000}x when applied to real-world networks.
At the same time the accuracy of the estimation is excellent.

Our contributions include experimentation on graphs with several millions of nodes and edges, where we show how practical our proposed method is. 
Finally, our algorithm's implementation is a part of the \pegasus library \footnote{Code and datasets are available at \url{http://www.cs.cmu.edu/~ctsourak/}.}
a Peta-Graph Mining library implemented in Hadoop, the open source version of Mapreduce. 

\end{abstract}

\section{Introduction}
\label{sec:intro}

Graphs are ubiquitous: the Internet, the World Wide Web (WWW), social networks, protein interaction networks 
and many other complicated structures are modelled as graphs. 
The problem of counting subgraphs is one of the typical graph mining tasks that has attracted 
a lot of attention (\cite{gspan}, \cite{closegraph}, \cite{yan2004} ) due to the wealth of applications related to it.
Indicatively we report the following:
a) Frequent small subgraphs are considered as a ``basis'', i.e., building blocks, for 
constructing classes of real-world networks \cite{milo:triangles},\cite{arena:com}.
b) In complex network analysis, computation of the transitivity ratio and the clustering 
coefficients requires computing the number of triangles in the graph \cite{newman:structure}. 
c) Community detection is a significant problem in many different scientific fields, e.g., parallel computation,
computer vision(\cite{ncut}), linear algebra(\cite{metis}), including graph mining \cite{leskovec,myra,metis}. 
Subgraph patterns such as bipartite cores or nearly ``bipartite cliques'', are used to detect emerging communities in the WebGraph \cite{kumar}
d) Fraudsters in online auction networks reportedly \cite{polo} seem to form specific patterns of connections, e.g., dense bipartite subgraphs.

The most basic, non-trivial subgraph, is the triangle. More formally, 
given a simple, undirected graph $G(V,E)$, a triangle is a three node fully connected subgraph. 
Many social networks have abundant triangles, since typically friends
of friends tend to become friends themselves \cite{faust:social}. This phenomenon is observed
in other types of networks as well (biological, online networks etc.) and is one of the main factors that 
gave rise to the definitions of the transitivity ratio  and the clustering coefficients of a graph \cite{newman:structure}.
Triangles have also been used in several applications. Namely, they have been used 
by Eckmann and Moses in \cite{eckman:thematic} to uncover the hidden thematic structure of the web and as a feature
to assist the classification of web activity as spamming or not, 
by Becchetti, Boldi, Castillo and Gionis in \cite{gionis:spam}.

In this paper we analyze a recent sampling algorithm for counting triangles
which appeared in \cite{Tsourakakiskdd09}. In \cite{Tsourakakiskdd09} 
only constant values of the sparsification parameter, i.e., sampling rate, 
were tested. A natural question to ask is how small can the sample be? 
If $p$ could be for example $O(\frac{1}{\sqrt{n}})$ while having guarantees that the estimate
is concentrated around the true value of the number of triangles in $G$, then 
the speedup would grow linearly with the number of nodes using an algorithm as the node iterator \cite{Tsourakakiskdd09}, 
giving tremendous speedups. 
Our main contribution is the rigorous analysis of Doulion \cite{Tsourakakiskdd09}, which yields optimal values
for the sparsification parameter $p$. We run our proposed method on large networks, 
showing speedups that reach the scale of about \emph{70000} faster performance with respect to the triangle counting task. 

The paper is organized as follows: Section~\ref{sec:prelim} presents briefly
the existing work and the theoretical background, Section~\ref{sec:method} 
presents our proposed optimal sampling method and Section~\ref{sec:experiments} presents
the experimental results on several large graphs. Section~\ref{sec:ramifications} presents 
two theoretical ramifications and in Section~\ref{sec:concl} we conclude.

\section{Preliminaries}
\label{sec:prelim}

In this section, we briefly present the existing work on the triangle counting problem 
and the necessary theoretical background of our analysis. \hide{We end this section by briefly 
describing Hadoop, our implementation platform.} Table \ref{tab:Symbols} lists the symbols used in this paper.

\begin{table}[t]
	\centering
	\begin{tabular}{|c|c|} \hline \hline
	Symbol & Definition \\ \hline \hline
  $G$ & a graph \\ \hline
  $n$ & number of nodes in $G$ \\ \hline
  $m$ & number of edges in $G$ \\ \hline
  $t$ & number of triangles in $G$ \\ \hline
  $\Delta(e)$ & $\#$ triangles  \\ \hline
              & that edge $e$ participates \\ 
	$\Delta$ & $max{\Delta(e)}$ \\ \hline
	$p$  & sparsification parameter \\ \hline
	$p^{*}$ & a $p$ value which gives  
	          concentration \\ \hline
	$p_I^{*}$ & ideal $p$ value, $p_I^{*}$=$min(p^*)$       \\    \hline
	$T$  & random variable, \\
	     & estimate of $t$\\ \hline \hline
\end{tabular}
	\caption{Table of symbols}
\label{tab:Symbols}
\end{table}

\subsection{Existing work}
There exist two general categories of triangle counting algorithms, the exact
and the approximating counting algorithms. 

\paragraph{Exact Counting}
The fastest exact counting methods use matrix-matrix multiplication and therefore
the overall time complexity is $O(n^{2.371})$,  which is the state of the art complexity
for matrix multiplication \cite{CopperWino:CopperWino}. The space complexity is $O(n^2)$. 
This category of algorithms are not used in practice due to the high memory requirements.
Even for medium sized networks, matrix-multiplication based algorithms are not applicable. 

Listing algorithms, even if they solve a more general problem than the counting one,
are preferred in practice for large graphs, due to the smaller memory requirements.
Simple representative algorithms are the node- and the edge-iterator algorithms.
In the former, at each iteration the algorithm considers the neighborhood of each node and counts
the number of edges among the neighbors, whereas the latter at each iteration considers and edge
and counts the common neighbors of the endpoints. Both have the same asymptotic complexity $O(mn)$, 
which in dense graphs results in $O(n^3)$ time, the complexity of the naive counting algorithm. 
Practical improvements over this family of algorithms have been achieved using various techniques, such as 
hashing to check if two nodes are neighborhood or not in constant time or sorting by the degree
to avoid unnecessary comparisons of neighborhoods of nodes (\cite{latapy,wagner:wagner}).

In planar graphs, Itai and Rodeh \cite{itai:rodeh} and also Papadimitriou and Yannakakis \cite{pap:yan} showed that triangles can be found in $O(n)$ time.
Itai and Rodeh in \cite{itai:rodeh} proposed an algorithm to find a triangle in any graph in $O(m^\frac{3}{2})$,
which can be extended to list the triangles in the graph with the same time complexity.
Their algorithm iteratively computes a spanning tree of the graph until there are no edges left, 
checks for each edge $(u,w)$ that does not belong to the spanning tree whether it belongs to
a triangle w.r.t the spanning tree and then removes the edges of the spanning tree.

The state of the art counting algorithm is due to Alon, Yuster and Zwick in \cite{alon:alon} and runs in $O(m^{\frac{2\omega}{\omega+1}})$,
where $\omega$=2.371, the fast matrix multiplication
exponent (\cite{CopperWino:CopperWino}). Thus, the Alon et al. algorithm currently runs in $O(m^{1.41})$ time. 

\paragraph{Approximate Counting} 
In many applications such as the ones mentioned in Section~\ref{sec:intro} the exact number of triangles is not crucial.
Thus approximating algorithms that are faster and output a high quality estimate are desirable. 
Most of the approximate triangle counting algorithms have been developed in the 
streaming setting. In this scenario, the graph is represented as a stream. 
Two main representations of a graph as a stream are the edge stream and the incidence stream. In the former, edges are arriving
one at a time. In the latter scenario
all edges incident to the same vertex appear successively in the stream. The ordering of the vertices 
is assumed to be arbitrary. A streaming algorithm produces a relative $\epsilon$ approximation 
of the number of triangles with high probability, making a constant number of passes over the stream. 
However, sampling algorithms developed in the streaming literature can be applied in the setting where the graph fits in the memory as well. 

Monte Carlo sampling techniques have been proposed to give a fast estimate of the number of triangles.
According to such an approach, a.k.a. naive sampling, we choose three nodes at random repetitively and check if they form a triangle or not. 
If one makes $$ r = \log({\frac{1}{\delta}})\frac{1}{\epsilon^2}(1+\frac{T_0+T_1+T_2}{T_3})$$
independent trials where $T_i=\#$triples with $i$ edges  and outputs as the estimate of triangles the random variable $ T_3' = {n \choose 3} \frac{\sum_{i=1}^r X_i}{r}$
then  $$(1-\epsilon)T_3 < T_3' < (1+\epsilon)T_3 $$ with probability at least $1- \delta$.
For graphs that have $T_3=o(n^2)$ triangles this approach is not suitable. This is the typical case, when dealing with real-world networks. 
This sampling approach is presented in \cite{shank:wanger1}.

Yosseff, Kumar and Sivakumar in their seminal paper \cite{yosseff} reduce the problem of triangle counting efficiently to estimating
moments for a stream of node triples. Then they use the Alon-Matias-Szegedy algorithms \cite{amsalgos} (a.k.a. AMS algorithms) to proceed. 
The key is that the triangle computation reduces in estimating the zero-th, first and second frequency moments, which can be done efficiently. 
Again, as in the naive sampling, the denser the graph the better the approximation.
The AMS algorithms are also used by \cite{jowhary}, where simple sampling techniques are used, such
as choose an edge from the stream at random and check how many common neighbors its two endpoints share considering the subsequent edges 
in the stream. 
In the same lines, Buriol et al. in \cite{buriol} proposed two space-bounded sampling algorithms to estimate the number of triangles. 
Again, the underlying sampling procedures are simple. E.g., for the case of the edge stream representation, they sample randomly
an edge and a node in the stream and check if they form a triangle. Their algorithms are the state-of-the-art algorithms to 
the best of our knowledge. In their three-pass algorithm, in the first pass they count the number of edges, in the second pass 
they sample uniformly at random an edge $(i,j)$ and a node $k \in V-\{i,j\}$ and in the third pass they test
whether the edges $(i,k),(k,j)$ are present in the stream. The number of draws that have to be done in order to get 
concentration (of course these draws are done in parallel), is of the order
$$ r = \log({\frac{1}{\delta}})\frac{2}{\epsilon^2}(3+\frac{T_1+2T_2}{T_3})$$
Even if the term $T_0$ is missing compared to the naive sampling, the graph has still to be fairly dense with respect
to the number of triangles in order to get an $\epsilon$ approximation with high probability. 

In the special case of ``power-law'' networks Tsourakakis \cite{me1} showed that the spectral counting of triangles 
can be efficient due to the spectrum properties of this category networks. This algorithm can be viewed as a special case
of a streaming algorithm, since there exist algorithms (\cite{tamas}) that perform a constant number of passes
over the non-zero elements of the matrix to make a good w.r.t the SVD, low rank approximation of a matrix.
In \cite{gionis:spam} the semi-streaming model for counting triangles is introduced. Becchetti et. al. observed
that since counting triangles reduces to computing the intersection of two sets, namely the induced neighborhoods
of two adjacent nodes, ideas from the locality sensitivity hashing \cite{alan} are applicable to the problem of counting
triangles. They relax the constraint of a constant number of passes over the edges, by allowing $\log n$ passes. 
 
\paragraph{DOULION} 

Doulion, a recent algorithm which appeared in \cite{Tsourakakiskdd09} proposed a new sampling procedure. 
The algorithm tosses a coin independently for each edge with probability $p$ to keep the edge and probability $q=1-p$
to throw it away. In case the edge ``survives'', it gets reweighted with weight equal to $\frac{1}{p}$.
\hide{Observe that since the initial graph was unweighted, all edges in the resulting sparsified graph $G'$ have
weight equal to $\frac{1}{p}$,  and thus a single number has to be stored, i.e., $p$. }
Then, any triangle counting algorithm, such as the node- or edge- iterator, is used to count
the number of triangles $t'$ in $G'$.  The estimate of the algorithm is the random variable $T=\frac{t'}{p^3}$.
The following facts  -among others- were shown in \cite{Tsourakakiskdd09}: 
\begin{itemize}
\item The estimator $T$ is unbiased, i.e., $\Mean T=t$.
\item The expected speedup when a simple exact counting algorithm as the node iterator is used, is $1/p^2$. 
\end{itemize}

The authors however did not answer a critical question: how small can $p$ be? In \cite{Tsourakakiskdd09}
constant factor speedups were obtained leaving the question as a topic of future research.

\subsection{Concentration of boolean Polynomials}
A common task in combinatorics is to show that if $Y$ is a polynomial
of independent boolean random variables then $Y$ is concentrated 
around its expected value. In the following we state the necessary definitions
and the main concentration result which we will use in our method. 

Let $Y=Y(t_1,\ldots,t_m)$ be a polynomial of $m$ real variables. The following definitions are from \cite{tao-vu}.
$Y$ is totally positive if all of its coefficients are non-negative variables,
regular if all of its coefficients are between zero and one, 
simplified if all of its monomials are square free and homogeneous if all of its monomials have the same degree. 
Given any multi-index $\alpha=(\alpha_1,\ldots,\alpha_m) \in \field{Z}^m_{+}$, define the partial derivative
${\partial^{\alpha}Y}= ( \frac{\partial}{\partial t_1} )^{\alpha_1} \ldots ( \frac{\partial}{\partial t_m} )^{\alpha_m} Y(t_1,\ldots,t_m)$
and denote by $\Abs{\alpha} = {\alpha_1}+\cdots{\alpha_m}$ the order of $\alpha$.
For any order $d \geq 0$, define ${\mathbb E}_d(Y)=\max_{\alpha:|\alpha|=d}{\mathbb E}(\partial^{\alpha}Y)$
and ${\mathbb E}_{\ge d}(Y) =\max_{d' \ge d}{\mathbb E}_{d'}(Y)$.

Typically, when $Y$ is smooth then it is strongly concentrated. 
By smoothness one usually means a small Lipschitz coefficient. In other words, when one changes the value of one 
variable $t_j$, the value $Y$ changes no more than a constant. However, as stated in \cite{vu} this is 
restrictive in many cases. Thus one can demand ``average smoothness'' as defined in \cite{vu}. 
For the purposes of this work, consider a random variable $Y=Y(t_1,\ldots,t_m)$ which is a positive polynomial 
of $m$ boolean variables $[t_i]_{i=1..m}$ which are independent. Observe that a boolean polynomial
is always regular and simplified.

Now, we refer to the main theorem of Kim and Vu of \cite[\S1.2]{kim-vu} as phrased in Theorem 1.1 of \cite{vu} or as Theorem 1.36\hide{\footnote{The notation $g(n)=O_k(f(n))$ means that there is a constant $d_k$ which depends on $k$ such that  $g(n) \leq d_k f(n)$ for all $n$.}} of \cite{tao-vu}.

\begin{theorem}
\label{thrm:kim-vu}
There is a constant $c_k$ depending on $k$ such that the following holds.
Let $Y(t_1, \ldots, t_m)$ be a totally positive polynomial of degree $k$, where $t_i$ can have arbitrary
distribution on the interval $[0, 1]$. Assume that:
\beql{cond1}
\Mean{Y} \ge {\mathbb E}_{\ge 1}(Y) 
\eeq
Then for any $\lambda \ge $ 1:
\beql{res1}
\Prob{\Abs{Y-\Mean{Y}} \ge c_k \lambda^k (\Mean{Y} {\mathbb E}_{\ge 1}(Y))^{1/2}} \le e^{-\lambda + (k-1)\log m}.
\eeq
\end{theorem}

\section{Proposed Method}
\label{sec:method}

\subsection{Analysis}

Now, we analyze a simple sparsification procedure which first appeared in \cite{Tsourakakiskdd09}:
toss a coin for each edge with probability $p$ to keep the edge and probability $q=1-p$
to throw it away. In case the edge ``survives'', we reweigh it with weight equal to $\frac{1}{p}$.
Observe that since the initial graph was unweighted, all edges in the resulting sparsified graph $G'$ have
weight equal to $\frac{1}{p}$, thus we just have to store a single number. Now, we count weighted triangles
in the sparsified graph $G'$. Our main result is the following theorem.

\begin{theorem}
\label{thrm:kolount}
Suppose $G$ is an undirected graph with $n$ vertices, $m$ edges and $t$ triangles. Let also $\Delta$ denote the size of the largest collection of triangles with a common edge. Let $G'$ be the random graph that arises from $G$ if we keep every edge with probability $p$ and write $T$ for the number of triangles of $G'$. Suppose that $\gamma>0$ is a constant and 
\beql{cond}
\frac{pt}{\Delta} \ge \log^{6+\gamma} n, \ \ \ \mbox{if $p^2\Delta \ge 1$},
\eeq
and
\beql{cond-small}
p^3 t \ge \log^{6+\gamma} n, \ \ \ \mbox{if $p^2\Delta < 1$}.
\eeq
for $n \ge n_0$ sufficiently large.
Then
$$
\Prob{\Abs{T-\Mean{T}} \ge \epsilon \Mean{T}} \le n^{-K}
$$
for any constants $K, \epsilon > 0$ and all large enough $n$ (depending on $K$, $\epsilon$ and $n_0$).
\end{theorem}

\begin{proof}
Write $X_e=1$ or $0$ depending on whether the edge $e$ of graph $G$ survives in $G'$. Then
$T = \sum_{\Delta(e,f,g)} X_e X_f X_g$ where
$\Delta(e,f,g) = \One{\mbox{edges $e,f,g$ form a triangle}}$.
Clearly $\Mean{T} = p^3 t$.

Refer to Theorem~\ref{thrm:kim-vu}. We use $T$ in place of $Y$, $k=3$.

We have
$$
\Mean{\frac{\partial T}{\partial X_e}} = \sum_{\Delta(e,f,g)} \Mean{X_f X_g} = p^2 \Abs{\Delta(e)},
$$
where $\Delta(e) = $ to how many triangles edge $e$ participates.
We first estimate the quantities ${\mathbb E}_j (T), j=0,1,2,3,$ defined before Theorem \ref{thrm:kim-vu}.
We get \beql{e1}
{\mathbb E}_1 (T) = p^2 \Delta
\eeq
where $\Delta = \max_{e} \Abs{\Delta(e)}$.

We also have
$$
\Mean{\frac{\partial^2 T}{\partial X_e \partial X_f}} = p\One{\exists g: \Delta(e,f,g)},
$$
hence
\beql{e2}
{\mathbb E}_2 (T) \le p.
\eeq
Obviously ${\mathbb E}_3(T) \le 1$.

Hence
$$
{\mathbb E}_{\ge 3} (T) \le 1,\ 
{\mathbb E}_{\ge 2} (T) \le 1,
$$
and
$$
{\mathbb E}_{\ge 1} (T) \le \max\Set{1, p^2\Delta},\ 
{\mathbb E}_{\ge 0} (T) \le \max\Set{1, p^2\Delta, p^3t}.
$$

\noindent
\underline{$\bullet$ {\sc Case 1} ($p^2\Delta < 1$):}\\
We get ${\mathbb E}_{\ge 1} (T) \le 1$, and, from \eqref{cond-small}, ${\mathbb E}_{\ge 0} (T) = p^3t$. 

\noindent \\
\underline{$\bullet$ {\sc Case 2} ($p^2\Delta \ge 1$):}\\
We get ${\mathbb E}_{\ge 1} (T) \le p^2\Delta$, and, from \eqref{cond}, ${\mathbb E}_{\ge 0} (T) = p^3t$.

We get, for some constant $c_3>0$, from Theorem \ref{thrm:kim-vu}:
\beql{dev}
\Prob{\Abs{T-\Mean{T}} \ge c_3 \lambda^3 (\Mean{T} {\mathbb E}_{\ge 1}(T))^{1/2}} \le e^{-\lambda + 2\log n}.
\eeq
Notice that in both cases we have $\Mean{T} \ge {\mathbb E}_{\ge 1}(T)$.

We now select $\lambda$ so that the lower bound inside the probability on the left-hand side of \eqref{dev} becomes $\epsilon\Mean{T}$.
In Case 1 we pick
$$
\lambda = \frac{\epsilon^{1/3}}{c_3^{1/3}} (p^3 t)^{1/6}
$$
while in Case 2
$$
\lambda = \frac{\epsilon^{1/3}}{c_3^{1/3}} \left(\frac{pt}{\Delta}\right)^{1/6}
$$
to get
\beql{dev1}
\Prob{\Abs{T-\Mean{T}} \ge \epsilon \Mean{T}} \le \exp(-\lambda+2\log n) 
\eeq
Since $\lambda \ge (K+2) \log n$ follows from our assumptions \eqref{cond} and \eqref{cond-small} if $n$ is sufficiently large, we get
$\Prob{\Abs{T-\Mean{T}} \ge \epsilon \Mean{T}} \le n^{-K}$, in both cases. 
\end{proof}

\subsection{Remarks}

This theorem states the important result that the estimator of the number of triangles is concentrated around its expected value, which 
is equal to the actual number of triangles $t$ in the graph \cite{Tsourakakiskdd09} under mild conditions on the triangle density of the
graph. The mildness comes from condition \eqref{cond}: picking $p=1$, given that our graph is not triangle-free, i.e., $\Delta \geq 1$,
gives that the number of triangles $t$ in the graph has to satisfy $t \geq \Delta \log^{6+\gamma} n$. 
This is a mild condition on $t$ since  $\Delta \leq n$ and thus it suffices that $t \geq n \log^{6+\gamma} n $ 
(after all, we can always add two dummy connected nodes that connect to every other node, as in Figure 1,
even if practically -experimentally speaking- $\Delta$ is smaller than $n$). 
The critical quantity besides the number of triangles $t$, is $\Delta$. Intuitively, if the sparsification
procedure throws away the common edge of many triangles, the triangles in the resulting graph may differ significantly
from the original. 

A significant problem is the choice of $p$ for the sparsification. The conditions \eqref{cond} and \eqref{cond-small} tell us how small we can afford to choose $p$, but the quantities involved, namely $t$ and $\Delta$, are unknown. One way around this obstacle would be to first estimate the order of magnitude of $t$ and $\Delta$ and then choose $p$ a little suboptimally. It may be possible to do this by running the algorithm a small number of times and deduce concentration if the results are close to each other. If they differ significantly then we sparsify less, say we double $p$, and so on, until we observe stability in our results. This would increase the running time by a small logarithmic factor at most. As we will describe in Section~\ref{sec:experiments}, in practice the doubling $p$ idea, works well. \hide{since concentration is achieved for small $p$ values, ranging from $0.005$ to $0.02$. Notice the corresponding expected speedups if one uses a simple counting algorithm like the node iterator \cite{Tsourakakiskdd09} are  2,500 to 40,000.}

From the theoretical point of view, this ambiguity of how to choose $p$ to be certain of concentration in our sparsification preprocessing does not however render our result useless. Under very general assumptions on the nature of the graph one should be able to get a decent value of $p$. For instance, if we we know $t \ge n^{3/2+\epsilon}$ and $\Delta\sim n$ , we get $p = n^{-1/2}$. This will result in a linear $O(n)$ expected speedup, as already mentioned in section~\ref{sec:prelim}.
On the other hand, if one wishes to make no assumptions on the nature of the graph, he/she can pick a constant $p$, e.g., $p=c$, and obtain expected speedups of order $\frac{1}{c^2}$ as described in \cite{Tsourakakiskdd09}.

\section{Experiments}
\label{sec:experiments}

In this section we describe first the experimental setup, and then we present the experimental results. 
We close the section by providing a practitioner's guide on how to use the analyzed triangle counting
algorithm through the detailed description of a specific experiment.

\subsection{Experimental Setup}

\paragraph{Datasets}
Table~\ref{tab:datasets} describes in brief the real-world networks we used in our experiments \footnote{Most of the datasets
can be found on the web, \url{http://www.cise.ufl.edu/research/sparse/matrices/}. The Youtube graph was made to us available 
upon request, \cite{mislove}}.
All graphs were first made undirected, and all self-loops were removed. The description 
of table~\ref{tab:datasets} refers to the graphs after the preprocessing. 

\paragraph{Algorithm}
We implemented the node iterator algorithm which was described in Section~\ref{sec:prelim} and was also used in \cite{Tsourakakiskdd09}. 
The code is written in JAVA and in Hadoop, the open source version of MapReduce \cite{dean}.

\paragraph{Machines}
We used two machines to run our experiments. The experiments for the three smallest graphs (Wikipedia 2005/9, 
Flickr, Youtube) were executed in a  2GB RAM, Intel(R) Core(TM)2 Duo CPU at 2.4GHz Ubuntu Linux machine.
For the three larger graphs (WB-EDU, Wikipedia 2006, Wikipedia 2005), we used the M45 supercomputer,
one of the fifty most powerful supercomputers in the world.
M45 has 480 hosts (each with  2 quad-core Intel Xeon 1.86 GHz, running
RHEL5),
with 3Tb aggregate RAM, and over 1.5 PetaByte aggregate disk capacity.
The cluster is running Hadoop on Demand (HOD). The number of machines allocated by HOD was set equal to three (3), 
given the relative small size of the graphs ($\approx$ 600-700 MB).
The sparsification triangle counting algorithm Doulion, i.e., sparsification and counting in the sparsified graph
were executed for all datasets in the Ubuntu machine. 
 
\begin{table}[ht]
\begin{center}
\begin{tabular}{|l|r|r|l|} \hline \hline
   Name  & Nodes & Edges &  Description  \\ \hline \hline
   WB-EDU&   9,845,725  &    46,236,105   &   Web Graph                    \\
                        &            &              & (page to page) \\ \hline
   Wikipedia  & 3,566,907 & 42,375,912  & Web Graph \\
   2007/2                    &            &              & (page to page) \\\hline
   Wikipedia & 2,983,494  & 35,048,116   & Web Graph  \\ 
    2006/6                    &            &              & (page to page) \\ \hline
   Wikipedia  & 1,634,989  & 18,540,603   & Web Graph   \\ 
    2005/9                   &            &              & (page to page) \\\hline
    Flickr & 404,733  & 2,110,078    & Person to Person \\ \hline
    Youtube &   1,157,822    &  4,945,382  & Person to Person\\ \hline
\end{tabular}
\end{center}
\caption{ Description of datasets }
\label{tab:datasets}
\end{table}

\subsection{Experimental Results}

Given that the majority of our datasets has $n$ of order $\approx 10^6$ we begin with a sparsification value $p=0.005$ which is $\approx 1/\sqrt{n}$.
We tried even smaller values than that (e.g, 0.001, 0.0005), but there was no concentration for any of the datasets.
We keep doubling the sparsification parameter until we deduce concentration and stop. In table~\ref{tab:results},
we report the results. In more detail, each row corresponds to the $p^{*}$ value, that we first deduced concentration using the doubling
procedure for each of the datasets we used (column 1). 
Ideally we would like to find	$p_I^{*}$, but we will settle with a $p^*$ value, since as already mentioned, doubling gives
at most an increase by a small logarithmic factor. 
Observe that $p^*$ is at most 2 times more than $p_I^*$ and upon its identification,
if one is curious about  $p_I^*$  for some reason, he/she can just do a simple ``binary'' search.
The third column of table~\ref{tab:results} described the quality of the estimator. Particularly, 
it contains values of the ratio \textit{our estimate}/\textit{\#triangles}. 
The next column contains the running time of the sparsification code, i.e., how much time it takes
to make one pass over the data and generate a second file containing the edges of the sparsified graph.
The fourth column \textbf{x}\textit{faster 1} contains the speedup of the node iterator vs. itself
when applied to the original graph and to the sparsified graph, i.e., the sample. 
The last column, \textbf{x}\textit{faster 2}, contains the speedup of the whole procedure we suggest, i.e., the doubling procedure,
counting and  repeat until concentration deduction, vs. running node iterator on the original graph. 

Some interesting points concerning these experimental results are the following:
a) The concentration we get is strong for small values of $p$, which implies directly large speedups.
b) The speedups typically are close to the expected ones, i.e., $\frac{1}{p^2}$ for the experiments
that we conducted in whole in the small (Ubuntu) machine. For the three experiments that were conducted
using Hadoop, the speedups were larger than the expected ones. This was (at least partially) expected since the necessary time for the JVM
(Java Virtual Machine) to load in M45, the disk I/O and most importantly the network communication
increase the running time for the node iterator algorithm when executed in parallel. 
However, for larger graphs that would span several Gigabytes, 
this speedup excession that we observed in our experiments should not show up as much.
The most important point to keep besides the system details is that our theorem guarantees concentration 
which implies that observing almost the same estimate in the sparsified graph multiple times
is equivalent to being able to make a good estimate for the true number of triangles. 
c) Even if the ``doubling-and-checking for concentration'' procedure may have to be repeated several times
the sparsification algorithm is still of high practical value. This is witnessed by the last column 
of the table. 
d) The overall speedups in the last column can easily be increased if one is willing to be less conservative
in the following sense: we conducted six experiments to deduct concentration. But in practice, one could conduct 
concentration using e.g., four experiments. Typically, concentration is easy to deduce. In the Wikipedia 2005/09
experiment, the first four experiments give 354, 349, 348 and 350 triangles in the sparsified graph which 
upon division with $0.02^3$ result in high accuracy estimates.
e) Finally, when concentration is deducted, averaging the concentrated estimates, typically gives a reasonable
estimator of high accuracy.

\begin{table}[ht]
\begin{center}
\begin{tabular}{c|c|c|c|c|c}\hline \hline 
       &         &     Mean Accuracy                &  Sparsify          &  x\it{faster}     &  x\it{faster}            \\ 
$G$    &$p^{*}$  &    ( 6 experiments)     &  (secs)            &  1                &     2                     \\ \hline \hline
WB     &0.005 & 95.8                            &  8                 & 70090             &  370.4                      \\
-EDU   &      &                                 &                    &                   &  														\\ 
       &      &                                 &                    &                   &                               \\ \hline
Wiki-  & 0.01 & 97.7                            &  17                & 21000             &  332                           \\
2007   &      &                                 &                    &                   &                                 \\ \hline
Wiki-  & 0.02 & 94.9                            &  14                & 4000              &  190.47                                \\
2006   &      &                                 &                    &                   &                                   \\ \hline
Wiki-  & 0.02 & 96.8                            &  8.6               & 2812              &  172.1                             \\
2005   &      &                                 &                    &                   &                                     \\ \hline
Flickr & 0.01 & 94.7                            &  1.2               & 12799             &  45                                  \\
       &      &                                 &                    &                   &                                       \\    \hline
You-   & 0.02 & 95.7                            &  2.3               & 2769              &  56                                    \\     
tube   &      &                                 &                    &                   &                                         \\ \hline \hline 
\end{tabular}
\end{center}
\caption{ Experimental results. Observe how small can $p$ be, resulting in huge savings during the triangle counting time. The 
``doubling-and-checking for concentration'' procedure that one would employ in practice gives important savings and high accuracy
at the same time. The drop-off in the total speedup is mainly due to the sparsification time.  }
\label{tab:results}
\end{table}

\begin{table}[ht]
\begin{center}
\begin{tabular}{c|c|c|c}
    &  & & Average   \\ 
     & Ratios & Sparsifi-   & Speedup  \\ 
    p    & \large{$\frac{T}{t}$ }   & cation (secs) & (x\it{faster})  \\ \hline \hline
   
   0.01 & 0.9442,  1.4499    &   8      &     7090               \\
        & 1.14, 1.37         &         &                    \\ \hline
   0.02 & 0.9112, 1.0183 & 8.64  & 2880 \\
        & 0.8975, 0.9579                                          &       & \\
        & 0.9716, 0.9771                                          &       & \\
        & 0.9524, 0.9551                                         &       & \\
        & 0.9606, 0.9716                                        &       & \\ \hline
   0.03 & 1.0043, 1.0336, 1.0035  &  8.65 &1500      \\
        & 0.9791, 1.0222 &    &   \\
        & 0.9865, 0.9816 &    &   \\  \hline
   0.04 & 0.9895, 1.018  &   8.58 &  825   \\ \hline
   0.05 & 0.9979, 0.9716    & 9.84 &  402   \\

\end{tabular}
\end{center}
\caption{ Wikipedia 2005/09: In this example, one deduces concentration for $p=0.02$. The corresponding speedup 
(node iterator on $G$ and on a small sample of $G$) averaged over the ten experiments is 2880 times. Results for $p$ greater
than 0.02 show that above that value strong concentration is achieved. }
\label{tab:wiki2005}
\end{table}

\subsection{A Practitioner's guide}

At first sight, according to theorem \ref{thrm:kolount} in order to pick the optimal value for 
$p$ we have to know the quantity that we are trying to compute, i.e., $t$ (and also $\Delta$). 
Even if one knows nothing about the triangle density of the graph of interest, or wishes
to make no assumptions, the proposed method is still of high practical value. 
In this subsection our goal is to provide a practitioner's guide. 
Specifically, we describe in detail how one can apply the sampling algorithm to 
a real world network, using our experimental experience as a guide, through an example.
Specifically, we describe how one can run the sampling algorithm in practice
by ``zooming'' in the Wikipedia 2005/9 experiment.

The Wikipedia 2005/9 graph after made undirected has $n=1,634,989$ nodes 
and $18,540,603$. The total number of triangles in the graph is $t=45,542,697$.
A simple computation gives that the triangle density $\frac{t}{{n \choose 3}}$ is 
equal to $6.25*10^{-11}$. This is a phenomenon that is observed with all the 
networks we used, i.e., very low triangle density. This should not be surprising, 
since ``real-world networks'' exhibit very skewed degree distributions. 
Roughly speaking, there exist many nodes with degree one, often connected to degrees of low
degree, e.g., 2. Immediately those nodes, i.e. nodes of degree 1 and of degree 2 that 
are connected with nodes of degree one, participate in no triangles. Furthermore, 
many nodes are totally disconnected, having degree zero. 
Even if the triangle density assumption  \eqref{cond} of our theorem is violated, 
the way to run the algorithm is the same. The value of $p$ will be necessarily bigger
to have concentration (the closer we get to a linear number of triangles, the larger $p$ gets
so as to have concentration), but as Table~\ref{tab:results} suggests, the method 
is of high practical value. 
One can start with a small sparsification value for $p=0.01$ \footnote{As described in the previous
subsection we start with even smaller value, but for brevity reasons, we begin here with 0.01 since concentration
appears for $p=0.02$.}

\begin{figure}
    \begin{center}
        \begin{tabular}{c}
        \includegraphics[width=0.45\textwidth]{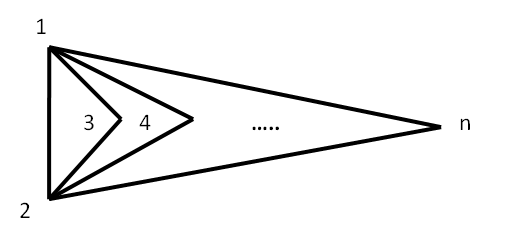} \\
	  \end{tabular}	    
		\caption{ Linear number of triangles: If our biased coin decides to delete edge $(1,2)$, then our sampling approach misses all triangles.  }
		\end{center}
	\label{fig:fig1}
\end{figure}

For $p=0.01$, running the sparsification code in a small machine 
with 2GB RAM, Intel(R) Core(TM)2 Duo CPU at 2.4GHz Ubuntu Linux machine, the sparsification
takes $~\approx 8$seconds and the counting (excluding the time to read the graph into the memory)
procedure using the simple node iterator algorithm takes 0.35 seconds. 
We ran this experiment four times, to make sure that this specific value of $p$ 
gives us the desired concentration. 
The number of triangles in the sparsified graph were found to be equal to 43, 66, 52 and 60.
Thus the estimates that the algorithm makes are respectively $4.3$x$10^7$,$6.6$x$10^7$,$5.2$x$10^7$ and $6$x$10^7$.
As one can observe, even if the average of those estimates gives an accuracy of 82.43\%,
the variance of those estimates is large. Thus, the value $p=0.01$ is not to be trusted. 
Doubling $p$, i.e., $p=0.02$, and running the code 10 times results in the following estimate of the number of triangles:
$4.15$x$10^7$, $4.25$x$10^7$, $4.6375$x$10^7$, $4.0875$x$10^7$, $4.3625$x$10^7$, $4.3375$x$10^7$, $4.35$x$10^7$, $4.375$x$10^7$, $4.45$x$10^7$, $4.25$x$10^7$.
The sparsification procedure takes $\approx 9$ seconds and the counting procedure in average $2.46$ seconds with variance equal to $0.1$ second
and can easily run in a machine with insufficient memory.
The speedup using $p=0.02$ due to our method is in average 2880, compared to running the node iterator in the initial graph. 
And as shown in the previous subsection the doubling idea still results in important speedups. 
If one tries slightly larger values for $p$, he/she would observe a strong concentration suggesting that we have a good estimate. 

The above are summarized in table~\ref{tab:wiki2005}. Each row corresponds to more than one experiments for a specific value of $p$. 
The first column shows the sparsification parameter, the second column contains the ratios $\frac{T}{t}$, and there are 
as many of them as the number of experiments were conducted for $p$ value in the same row, the third column
contains the running time of the sparsification procedure and the last column the average speedup obtained when 
we run the node iterator on the whole graph and on the small sample we obtain using the sparsification procedure. 
In this example, one can deduce at $p=0.02$ concentration and stop running the algorithm. 
As we observe, given a graph $G$ the sparsification time is more or less the same (8-9sec), correlated positively
with $p$, as more I/O write operations are being done (writing edges to a new file). The speedup we get averaged
over the experiments we did compared to the expected one, can be approximately 
the (e.g., $p=$0.05), can be larger (e.g., $p=$0.02) and can be also smaller (e.g., $p=$0.01).

\section{Theoretical Ramifications} 
\label{sec:ramifications}
\subsection{Linear number of triangles}

One may wonder how the algorithm performs in graphs where the number of triangles is linear, i.e., $O(n)$.
Consider the graph of figure 1. If the coin decides that the common
edge should be removed then we lose all the triangles. Thus the sparsification step may introduce 
an arbitrarily high error in our estimate.

\subsection{Weighted graphs}

\begin{figure}
    \begin{center}
        \begin{tabular}{c}
		    \includegraphics[width=0.45\textwidth]{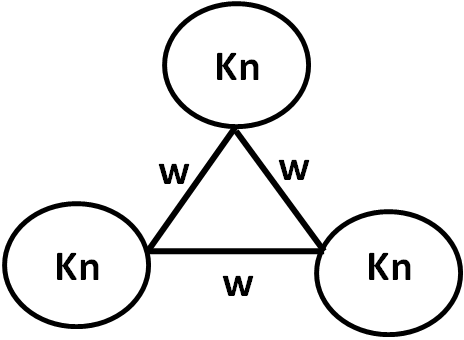} \\
	  \end{tabular}	    
		\caption{ Weighted case: For $w$ sufficiently large, our sampling approach can perform badly if one of the weighted edges gets deleted. }
		\end{center}
	\label{fig:fig2}
\end{figure}

Consider now the case of weighted graphs. The algorithm of \cite{Tsourakakiskdd09}
can be extended to weighted graphs: each edge gets reweighted with weight equal to 
the old weight times $1/p$. 
However, one can come up with counterexamples that show that this algorithm 
can perform badly on weighted graphs.
Such an example where the algorithm can perform badly is shown in figure 2.
If $w$ is large enough, then the removal of one of the weighted edges will introduce 
a large error in the final estimate. 

\section{Conclusions}
\label{sec:concl}
We present an algorithm that under mild conditions on the triangle density of the graph
performs accurately, i.e., outputs a good estimate of the number of triangles, with high probability. 

Our main contributions are:
\begin{itemize}
\item The analysis of the sparsification algorithm, which leads to optimal values of the 
sparsification parameter $p$. Thus, we can justify speedups rigorously rather than the constant speedups of  \cite{Tsourakakiskdd09}.
\item A practitioner's guide on how to run the algorithm in detail. Even if the optimal values
of $p$ depend on unknown quantities, including the number of triangles we wish to estimate, 
the algorithm is of high practical value. Few executions until concentration is deduced, still result 
in huge speedups. 
\item Experimentation on large networks, with several millions of nodes and edges.
\end{itemize}

Finally, both cases presented in Section~\ref{sec:ramifications} require a sophisticated sampling procedure (e.g., \cite{spielman:spielman}), 
rather than a simple one and these are topics of future research. 

\section{Acknowledgments}
The first author would like to thank Alan Frieze, Petros Drineas and Ioannis Koutis for helpful discussions. 
This material is based upon work supported by the National Science Foundation under Grants No. CCF-0635257,
No. IIS-0705359 and by the University of Crete under Grant No. 2569.

%
%


\begin{thebibliography}{5}

\bibitem{amsalgos}
Alon, N., Yossi, M., Szegedy, M.:
The space complexity of approximating the frequency moments
Proceedings of ACM STOC (1996)

\bibitem{alon:alon}
Alon, N., Yuster, R., Zwick, U.:
Finding and Counting Given Length Cycles.
In Algorithmica, Volume 17, Number 3, 209--223 (1997)

\bibitem {arena:com}
Arenas, A., Fernandez, A., Fortunato, S., Gomez, S.:
Motif-based communities in complex networks
J. Phys. A: Math. Theor. (2008)


\bibitem{yosseff}
Bar-Yosseff, Z., Kumar, R., Sivakumar, D.:
Reductions in streaming algorithms, with an application to counting triangles in graphs.
Proceeding of ACM-SIAM SODA (2002)

\bibitem {gionis:spam}
Becchetti, L., Boldi, P., Castillo, C.,  Gionis, A.:
Efficient Semi-Streaming Algorithms for Local Triangle Counting in Massive Graphs.
Proceedings of ACM KDD, 2008.

\bibitem{alan}
Broder, A.Z., Charikar, M., Frieze, A., Mitzenmacher, M.:
Min-wise independent permutations.
Proceedings of ACM STOC 1998

\bibitem{buriol}
Buriol, L., Frahling, G., Leonardi, S.,  Marchetti-Spaccamela, A., Sohler, C.:
Counting Triangles in Data Streams
PODS 2006


\bibitem {CopperWino:CopperWino}
Coppersmith D., Winograd S.:
Matrix multiplication via arithmetic progressions.
Proceedings of ACM STOC (1987)

\bibitem{dean}
Jeffrey, D., Ghemawat, S.:
MapReduce: Simplified Data Processing on Large Clusters
OSDI '04

\bibitem {eckman:thematic}
Eckmann, J.-P., Moses, E.:
Curvature of co-links uncovers hidden thematic layers in the World Wide Web.
PNAS (200)








\bibitem{itai:rodeh}
Itai, A., Rodeh, M.:
Finding a minimum circuit in a graph.
Proceedings of ACM STOC (1977)


\bibitem{jowhary}
Jowhari, H., Ghodsi, M.:
New Streaming Algorithms for Counting Triangles in Graphs
COCOON 2005, 710--716

\bibitem{gspan}
Yan, X., Han, J.:
gSpan: Graph-Based Substructure Pattern Mining,
ICDM 2002

\bibitem{closegraph}
Yan, X., Han, J.:
CloseGraph: mining closed frequent graph patterns,
KDD '03: Proceedings of the ninth ACM SIGKDD international conference on Knowledge discovery and data mining


\bibitem{yan2004}
Yan, X., Yu, P., Han, J.:
Graph indexing: a frequent structure-based approach
SIGMOD '04: Proceedings of the 2004 ACM SIGMOD international conference on Management of data


\bibitem{kim-vu}
J.H. Kim and V.H. Vu,
Concentration of multivariate polynomials and its applications,
Combinatorica {\bf 20} (2000), 3, 417--434.


\bibitem{metis}
Karypis, G., Kumar, V.,
METIS - Unstructured Graph Partitioning and Sparse Matrix Ordering System, Version 2.0
1995

\bibitem{kumar}
Kumar, R., Raghavan, P., Rajagopalan, S., Tomkins, A.:
Trawling the Web for emerging cyber-communities
Computer Networks (Amsterdam, Netherlands: 1999)

\bibitem{latapy}
Latapy, M.:
Main-memory triangle computations for very large (sparse (power-law)) graphs
Theor. Comput. Sci., 407, 458--473 (2008)

\bibitem{leskovec}
Leskovec, J., Lang, K.,  Dasgupta, A., Mahoney, M.:
Statistical properties of community structure in large social and information networks
WWW '08: Proceeding of the 17th international conference on World Wide Web

\bibitem {milo:triangles}
Milo, R., Shen-Orr, S., Itzkovitz, S., Kashtan, N., Chklovskii, D., Alon, U.:
Network motifs: simple building blocks of complex networks.
Science (2002)

\bibitem{mislove}
Mislove, A., Massimiliano, M., Gummadi, K., Druschel, P., Bhattacharjee, B.:
Measurement and Analysis of Online Social Networks
Proceedings of the 5th ACM/Usenix Internet Measurement Conference (IMC'07)

\bibitem {newman:structure}
Newman, M.:
The structure and function of complex networks (2003).


\bibitem{pap:yan}
Papadimitriou, C., Yannakakis, M.:
The clique problem for planar graphs. 
Information Processing Letters, 13, 131--133 (1981).


\bibitem{polo}
Pandit, Shashank, Chau, Duen  H., Wang, S., Faloutsos, C.:
Netprobe: a fast and scalable system for fraud detection in online auction networks
WWW '07: Proceedings of the 16th international conference on World Wide Web



\bibitem{ncut}
Shi, J., Malik, J.:
Normalized cuts and image segmentation
Pattern Analysis and Machine Intelligence, 2000



\bibitem{tamas}
Sarlos, T.:
Improved Approximation Algorithms for Large Matrices via Random Projections
Proceedings of FOCS (2006)


\bibitem{myra}
Falkowski, T., Barth, A., Spiliopoulou,M.:
DENGRAPH: A Density-based Community Detection Algorithm
WI '07: Proceedings of the IEEE/WIC/ACM International Conference on Web Intelligence


\bibitem {wagner:wagner}
Schank, T., Wagner, D.: 
Finding, Counting and Listing all Triangles in Large Graphs, An Experimental Study
WEA (2005)

\bibitem{shank:wanger1}
Schank, T., Wagner, D.:
Approximating Clustering Coefficient and Transitivity.
Journal of Graph Algorithms and Applications, 9, 265--275 (2005)

\bibitem {spielman:spielman}
Spielman, D., Srivastava, N.:
Graph Sparsification by Effective Resistances.
Proceedings of ACM STOC (2008)


\bibitem{tao-vu}
T. Tao and V.H. Vu, {\em Additive Combinatorics},
Cambridge Univ.\ Press 2006.

\bibitem{me1}
Tsourakakis, C.E.:
Fast Counting of Triangles in Large Real Networks, without counting: Algorithms and Laws
ICDM 2008

\bibitem{Tsourakakiskdd09}
Tsourakakis, C.E., Kang, U, Miller, G.L., Faloutsos, C.:
Doulion: Counting Triangles in Massive Graphs with a Coin
Proceedings of ACM KDD, 2009


\bibitem{vu}
V.H. Vu,
On the concentration of multivariate polynomials with small expectation,
Random Structures and Algorithms {\bf 16} (2000), 4, 344--363.

\bibitem {faust:social}
Wasserman, S., Faust, K.:
Social Network Analysis : Methods and Applications (Structural Analysis in the Social Sciences).
Cambridge University Press (1994)


\end{thebibliography}
\end{document}